# Stuckelberg SUSY QED and Infrared Problem

**Radhika Vinze**$^{a,\$}$, **T. R. Govindarajan** $^{b,\dagger}$, **Anuradha Misra**$^{a,\$\$}$, **P. Ramadevi** $^{c,\dagger\dagger}$

$^a$*Department of Physics, University of Mumbai, Mumbai 400098, India.*

$^b$ *Chennai Mathematical Institute, Kelambakkam Siruseri, Tamil Nadu 600113, India.*

$^c$ *Department of Physics, Indian Institute of Technology Bombay, Powai, Mumbai 400076, India.*

*

## Abstract

We review gauge invariant $\mathcal{N} = 1$ supersymmetric massive $U(1)$ gauge theory coupled to matter and Stuckelberg superfields. We focus on the leading order self energy and vertex correction to the matter field in the massless limit of both the $U(1)$ vector superfield and the Stuckelberg superfield. We explicitly verify that the theory is infrared divergence free in the massless limit. Hence the Stuckelberg mechanism appears to be the efficient route to handle infrared divergences seen in supersymmetric quantum electrodynamics. Since these additional particles have very small masses they can serve as dark matter candidates through 'Ultralight particles' mechanism.

* $^\$$ radhika.vinze@physics.mu.ac.in, $^\dagger$ trg@cmi.ac.in, trg@imsc.res.in, $^{\$\$}$ misra@physics.mu.ac.in,

$^{\dagger\dagger}$ramadevi@phy.iitb.ac.in




# I. INTRODUCTION

Quantum Electrodynamics (QED) is a classic example of a theory with infrared (IR) divergences which are encountered in theories with a massless gauge boson. Conventional way to deal with these divergences, which arise due to small energy of a virtual photon in a loop diagram, is to take into account the emission of real soft photons. This approach based on real-virtual cancellation is a cross section approach[1, 2]. An alternative method, where the cancellation of IR divergences takes place at amplitude level itself, is the coherent state approach[3–5]. This has been applied to QED as well as other theories with massless particles to remove the infrared divergences. The coherent state approach has been developed in light-front field theory[6, 7] and has been shown to lead to cancellation of IR divergences in LFQED[7, 8].

An alternative approach to deal with IR divergences is through the Stuckelberg mechanism. A massive $U(1)$ gauge theory coupled to matter field can be made gauge invariant with the introduction of an additional massive scalar field called Stuckelberg field[9]. This is known in the literature as *Stuckelberg mechanism* of giving mass to vector boson without breaking gauge invariance. In fact, the mass of the gauge boson makes the theory IR finite. Further such a massive QED with the Stuckelberg field is a renormalisable theory [10, 11]. It is important to ascertain that the theory continues to be gauge invariant and IR finite even in the massless limit of the gauge boson and the Stuckelberg field. Indeed this was verified in the recent paper[12] by showing that the leading order self-energy correction and vertex correction are free from IR divergence. In that paper [12], a light-front Hamiltonan approach to QED was used to show the cancellation of IR divergences in a manner similar to the coherent state method of Ref.[7, 8]. Besides QED, the Stuckelberg mechanism has been applied on the abelian subgroup $U(1)_X$ of the minimally extended standard model $(SU(3) \times SU(2)_L \times U(1)_Y \times U(1)_X)$[13].

Supersymmetry achieved its fame when it was found to be a natural mechanism to protect the mass scale of Higgs boson[14]-[19]. Further, supersymmetry (SUSY) appears to be an important factor in string theory too. But the problem of infrared divergences is present in SUSY theories because of the presence of massless vector particles. For instance, $\mathcal{N} = 1$ supersymmetric QED (SUSY QED) does have IR divergences due to massless vector superfield. Interestingly, the supersymmetric extension of the Stuckelberg mechanism of



QED was investigated and shown to be a renormalisable theory in Ref.[20]. Inclusion of Fayet-Iliopoulos term in such a theory indeed preserves supersymmetry[21]. Further, supersymmetric Stuckelberg $U(1)$ extension of minimal supersymmetric standard model (MSSM) has also been studied in Refs. [22]. It is not a priori clear whether the massless limit of the vector superfield in the Stuckelberg SUSY QED is also IR finite. Hence, in this paper we attempt a similar exercise[12] for the Stuckelberg SUSY QED. Particularly, we explicitly work out the one loop self-energy diagrams as well as the vertex correction diagrams of matter for the Stuckelberg SUSY QED in the massless limit of both vector superfield and the chiral Stuckelberg superfield. Even though we study Stuckelberg theory with SUSY with the motivation to resolve infrared divergences in SUSY theories, we would like to point out other interesting features of the theory. Recall, the Stuckelberg field with same mass $M$ as photon was initially introduced mainly to preserve gauge invariance in the massive gauge theory. Taking the experimental bound on the photon mass $M \leq \mathcal{O}(10^{-18})eV$ [24]−[27], the electromagnetic interaction of the longitudinal/Stuckelberg particles with matter are weak[28]. Important realisation is the conserved charge excites longitudinal mode only at $\mathcal{O}(M^2)$ as shown in [29]. As the ultralight mass $M$ Stuckelberg field can also have gravitational interaction, the dark matter question can be addressed within the Stuckelberg QED approach [30]. Notice that there will be now two candidates in the Stuckelberg SUSY QED- namely, Stuckelberg field and its fermionic superpartner. Hence, the extension of SUSY QED with Stueckelberg field may be a direction to pursue in the hunt for dark matter candidates which are ultralight boson or axion like [31, 32]. Further, Dvali et al [33] have considered the use of Stuckelberg degrees of freedom playing a role in holography which strengthens our interest on the Stuckelberg SUSY QED.

The plan of the paper is as follows: In section II, we will first recapitulate the superfield expansion notation and then review the Lagrangian describing Stuckelberg SUSY QED. Focussing on the interaction terms present in such a Lagrangian, we perform the leading order contribution to the self-energy of the electron in section III and the vertex correction in section IV. Finally, we summarize our results in the concluding section V.



## II. STUCKELBERG SUSY QED

We can consider SUSY QED with a single massless chiral field. But this will have two problems:

- There will be additional divergence due to collinear momenta.
- Since $TrU(1) \neq 0$ there will be additional anomaly [34].

To avoid these problems, we will consider massive matter multiplet. The electron in the conventional QED is represented by a Dirac spinor. In the supersymmetric context, we work with Weyl spinors. So, we require two Weyl spinors which combine to give the Dirac field of the electron. Hence, the matter field Lagrangian will involve two chiral superfields $\Phi_+, \Phi_-$. The Lagrangian representing $U(1)$ gauge invariant $\mathcal{N}=1$ supersymmetric QED (SQED) is

$$\mathcal{L}_{SQED} = \mathcal{L}_V + \mathcal{L}_+ + \mathcal{L}_- + \mathcal{L}_m . \tag{1}$$

Here, $\mathcal{L}_V$ is the vector superfield term, $\mathcal{L}_\pm$ gives the interaction of $V$ with the charged matter fields $\Phi_\pm$ whose charge $\pm q$ and $\mathcal{L}_m$ denotes the mass term of $\Phi_+, \Phi_-$:

$$\begin{aligned}
\mathcal{L}_V &= \frac{1}{2}\left(W^\alpha W_\alpha + \text{h.c}\right) , \\
\mathcal{L}_\pm &= \Phi_\pm^\dagger e^{\pm 2qV} \Phi_\pm , \\
\mathcal{L}_m &= m(\Phi_+ \Phi_- + \Phi_+^\dagger \Phi_-^\dagger) ,
\end{aligned} \tag{2}$$

where $W^\alpha = \bar{D}^2 D^\alpha V$. We will take $V$ to be in the Wess-Zumino gauge. In this gauge, the component expansion of $V$ will be

$$V = \theta \sigma^\mu \bar{\theta} A_\mu + i\theta\theta\bar{\theta}\bar{\lambda} - i\bar{\theta}\bar{\theta}\theta\lambda + \frac{1}{2}\theta\theta\bar{\theta}\bar{\theta} D , \tag{3}$$

where $A_\mu$ is the gauge field and $\lambda$ is the gaugino Weyl spinor and $D$ is the auxillary field. The explicit form of $\mathcal{L}_V$ will be

$$\mathcal{L}_V = -\frac{1}{4} F^{\mu\nu} F_{\mu\nu} - i\lambda \sigma^\mu \partial_\mu \bar{\lambda} + \frac{1}{2} D^2 , \tag{4}$$

where $F_{\mu\nu} = \partial_\mu A_\nu - \partial_\nu A_\mu$ is the electromagnetic field strength. It is easy to check that under any gauge transformation

$$V \to V + i(\Lambda - \Lambda^\dagger) \quad ; \quad \Phi_\pm \to e^{\pm 2iq\Lambda} \Phi_\pm ,$$



$\mathcal{L}_V$ and $\mathcal{L}_\pm$ are invariant. The explicit form $\mathcal{L}_\pm$ in the Wess-Zumino gauge can be simplified since we can truncate $e^{2qV}$ as

$$e^{2qV} = 1 + 2qV + 2q^2V^2 \ .$$

Recall $V^n$ for $n > 2$ have more than two powers of $\theta$ and $\bar{\theta}$ and hence contribute zero to the action. Therefore, the simplified form of $\mathcal{L}_\pm$ turns out to be

$$\mathcal{L}_\pm = \Phi_\pm^\dagger \Phi_\pm \pm 2q\Phi_\pm^\dagger V \Phi_\pm + 2q^2 \Phi_\pm^\dagger V^2 \Phi_\pm \ . \tag{5}$$

Using the following component expansion of the chiral supermultiplet $\Phi(\phi_\pm, \psi_\pm, F_\pm)$

$$\Phi_\pm = \phi_\pm + \sqrt{2}\theta\psi_\pm + \theta\theta F_\pm + i\partial_\mu \phi_\pm \theta\sigma^\mu \bar{\theta} - \frac{i}{\sqrt{2}}\theta\theta \partial_\mu \psi_\pm \sigma^\mu \bar{\theta} - \frac{1}{4}\partial^2 \phi_\pm \theta\theta\bar{\theta}\bar{\theta} \ , \tag{6}$$

we can rewrite

$$\mathcal{L}_{SQED} = \mathcal{L}_V + \mathcal{L}_+ + \mathcal{L}_- + \mathcal{L}_m = \mathcal{L}_V + \mathcal{L}_{kin}^+ + \mathcal{L}_{kin}^- + \mathcal{L}_m + \mathcal{L}_{int}^+ + \mathcal{L}_{int}^- \ , \tag{7}$$

so that we can see them as kinetic terms of the matter fields and its interaction with vector superfield $V$:

$$\begin{aligned}
\mathcal{L}_{kin}^\pm &= \Phi_\pm^\dagger \Phi_\pm = |F_\pm|^2 + \partial^\mu \phi_\pm^* \partial_\mu \phi_\pm - i\bar{\psi}_\pm \bar{\sigma}^\mu \partial_\mu \psi_\pm \ , \\
\mathcal{L}_m &= +m(F_+^* \phi_-^* + F_-^* \phi_+^* + F_+ \phi_- + F_- \phi_+ + \bar{\psi}_- \bar{\psi}_+ + \psi_- \psi_+) \ , \\
\mathcal{L}_{int}^\pm &= \pm 2q\Phi_\pm^\dagger V \Phi_\pm + 2q^2 \Phi_\pm^\dagger V^2 \Phi_\pm \\
&= \pm q\left(iA^\mu \phi_\pm^* \partial_\mu \phi_\pm - iA^\mu \phi_\pm \partial_\mu \phi_\pm^* + A_\mu \psi_\pm \sigma^\mu \bar{\psi}_\pm + D\phi_\pm \phi_\pm^* \right. \\
&\quad \left. -\sqrt{2}i\phi_\pm \bar{\lambda}\bar{\psi}_\pm + \sqrt{2}i\lambda\psi_\pm \phi_\pm^*\right) + q^2 A^\mu A_\mu \phi_\pm \phi_\pm^* \ .
\end{aligned} \tag{8}$$

So far, we have elaborated $\mathcal{L}_{SQED}$ in terms of component fields which defines $\mathcal{N} = 1$ supersymmetric QED Lagrangian (SQED). The massless photons in SQED lead to infrared divergences. Just like the Stuckelberg mechanism in conventional QED removes IR divergence, we would like to add mass $M$ to the vector superfield and introduce Stuckelberg chiral superfield $S$ such that the Lagrangian $\mathcal{L}_M$ is gauge invariant. Thus under the gauge transformation of

$$V \to V + i(\Lambda - \Lambda^\dagger) \text{ and } S \to S - M\Lambda \ ,$$

the following $\mathcal{L}_M$ is gauge invariant:

$$\mathcal{L}_M = M^2(V')^2 = M^2[V + \frac{i}{M}(S - S^\dagger)]^2 \ . \tag{9}$$



Recall $V$ is in Wess-Zumino gauge but not $V'$. Just like the $\Phi_\pm$ superfield expansion, we can do component field expansion of the the Stuckelberg chiral superfield $S$:

$$S = \beta + \sqrt{2}\theta\tau + \theta\theta C + i\partial_\mu\beta\theta\sigma^\mu\bar{\theta} - \frac{i}{\sqrt{2}}\theta\theta\partial_\mu\tau\sigma^\mu\bar{\theta} - \frac{1}{4}\partial^2\beta\theta\theta\bar{\theta}\bar{\theta} \ . \tag{10}$$

Using the above form and its conjugate $S^\dagger$, we obtain

$$\begin{aligned}S - S^\dagger &= \beta - \beta^* + \sqrt{2}(\theta\tau - \bar{\theta}\bar{\tau}) + (\theta\theta C - \bar{\theta}\bar{\theta}C^*) + i\partial_\mu(\beta + \beta^*)\theta\sigma^\mu\bar{\theta} \\ &\quad - \frac{i}{\sqrt{2}}\theta\theta\partial_\mu\tau\sigma^\mu\bar{\theta} - \frac{i}{\sqrt{2}}\theta\partial_\mu\bar{\tau}\sigma^\mu\bar{\theta}\bar{\theta} - \frac{1}{4}\partial^2(\beta - \beta^*)\theta\theta\bar{\theta}\bar{\theta} \ .\end{aligned} \tag{11}$$

Let us take complex field $\beta = \dfrac{X + iY}{2}$ where $X, Y$ are real scalar fields. Then the above equation can be simplified as

$$\begin{aligned}S - S^\dagger &= [iY + \sqrt{2}(\theta\tau - \bar{\theta}\bar{\tau}) + (\theta\theta C - \bar{\theta}\bar{\theta}C^*) + i\partial_\mu X\theta\sigma^\mu\bar{\theta} \\ &\quad - \frac{i}{\sqrt{2}}\theta\theta\partial_\mu\tau\sigma^\mu\bar{\theta} - \frac{i}{\sqrt{2}}\theta\partial_\mu\bar{\tau}\sigma^\mu\bar{\theta}\bar{\theta} - \frac{i}{4}\partial^2 Y\theta\theta\bar{\theta}\bar{\theta}] \ .\end{aligned} \tag{12}$$

With the above component expansion, the vector superfield $V'$ is clearly in the non-Wess Zumino gauge:

$$\begin{aligned}V' &= -\frac{Y}{M} + \frac{\sqrt{2}i}{M}(\theta\tau - \bar{\theta}\bar{\tau}) + \frac{i}{M}(\theta\theta C - \bar{\theta}\bar{\theta}\bar{C}) + \theta\sigma^\mu\bar{\theta}\,\overbrace{(A_\mu - \frac{1}{M}\partial_\mu X)}^{\tilde{A}_\mu} \\ &\quad + i\theta\theta\bar{\theta}(\bar{\lambda} - \frac{i}{\sqrt{2}M}\partial_\mu\tau\sigma^\mu) - i\bar{\theta}\bar{\theta}\theta(\lambda + \frac{i}{\sqrt{2}M}\partial_\mu\bar{\tau}\sigma^\mu) \\ &\quad + \frac{1}{2}\theta\theta\bar{\theta}\bar{\theta}(D + \frac{1}{2M}\partial^2 Y) \ .\end{aligned} \tag{13}$$

Notice the gauge field $A_\mu$ comes along with $\partial_\mu X$. We can make a change of variable

$$\tilde{A}_\mu = A_\mu - \frac{1}{M}\partial_\mu X \ . \tag{14}$$

Remember, we have to replace $A_\mu$ in $\mathcal{L}_{int}^\pm$ as $\tilde{A}_\mu + \frac{1}{M}\partial_\mu X$ so that the total Lagrangian involves $\tilde{A}_\mu$. The explicit form of Lagrangian $\mathcal{L}_M$ in terms of $\tilde{A}_\mu$ is

$$\begin{aligned}\mathcal{L}_M &= \frac{1}{2}M^2\tilde{A}^\mu\tilde{A}_\mu - MYD + \frac{1}{2}\partial^\mu Y\partial_\mu Y + 2CC^* + 2i\tau\sigma^\mu\partial_\mu\bar{\tau} \\ &\quad - \sqrt{2}M(\bar{\lambda}\bar{\tau} + \lambda\tau) \ .\end{aligned} \tag{15}$$

Using the equations of motion for the auxiliary fields $F_\pm, C, D$:

$$F_+ = -m\phi_-^*, \ F_- = -m\phi_+^*, \ F_+^* = -m\phi_-, \ F_-^* = -m\phi_+,$$
$$C = 0, \ D = MY + q|\phi_-|^2 - q|\phi_+|^2 \ ,$$



we eliminate these fields in the total Lagrangian $\mathcal{L} = \mathcal{L}_{SQED} + \mathcal{L}_M$. The final form of $\mathcal{L}$ is

$$\mathcal{L} = -\frac{1}{4}F^{\mu\nu}F_{\mu\nu} + \frac{1}{2}M^2\tilde{A}^\mu\tilde{A}_\mu - \frac{1}{2}q^2(|\phi_+|^2 - |\phi_-|^2)^2 + \sum_{j=+,-}\partial^\mu\phi_j\partial_\mu\phi_j^* - m^2|\phi_j|^2$$

$$+iq(\tilde{A}_\mu + \frac{1}{M}\partial_\mu X)\sum_{j=+,-}\left(j[\phi_j^*\partial^\mu\phi_j - \phi_j\partial^\mu\phi_j^*] - iq(\tilde{A}^\mu + \frac{1}{M}\partial^\mu X)|\phi_j|^2\right)$$

$$+\sum_{j=+,-}\left(i\psi_j\sigma^\mu\partial_\mu\bar{\psi}_j + jq\bar{\psi}_j\sigma^\mu(\tilde{A}_\mu + \frac{1}{M}\partial_\mu X)\psi_j\right) - m(\bar{\psi}_-\bar{\psi}_+ + \psi_-\psi_+)$$

$$+i\lambda\sigma^\mu\partial_\mu\bar{\lambda} - \sqrt{2}M(\bar{\lambda}\bar{\tau} + \lambda\tau) + 2i\tau\sigma^\mu\partial_\mu\bar{\tau} + \sqrt{2}iq\bar{\lambda}(\phi_-\bar{\psi}_- - \phi_+\bar{\psi}_+)$$

$$+\sqrt{2}iq\lambda(\psi_+\phi_+^* - \psi_-\phi_-^*) + \frac{1}{2}\partial_\mu Y\partial^\mu Y - \frac{1}{2}M^2 Y^2 + qMY(|\phi_+|^2 - |\phi_-|^2) \ . \quad (16)$$

We now introduce the following gauge fixing term to the above Lagrangian:

$$\mathcal{L}_{gf} = -\frac{1}{2\alpha}(\partial_\mu A^\mu + \alpha MX)^2 = -\frac{1}{2\alpha}(\partial_\mu\tilde{A}^\mu + \frac{1}{M}\Box X + \alpha MX)^2 \ . \quad (17)$$

In Feynman gauge $\alpha = 1$ and using equation of motion for Stuckelberg scalar field $X$:

$$\Box X + M^2 X = 0 \ ,$$

the gauge fixing term is simplified to

$$\mathcal{L}_{gf} = -\frac{1}{2}(\partial_\mu\tilde{A}^\mu)^2 + \frac{1}{2}\partial^\mu X\partial_\mu X - \frac{1}{2}M^2 X^2 \ . \quad (18)$$

Now we introduce a Dirac spinor $\Psi_E$ for the matter field Weyl spinors $\psi_+, \psi_-$ and another Dirac spinor $\Psi_G$ for gaugino $\lambda$ and the Stuckelberg superpartner (Weyl spinor) $\tau$. We also introduce charge conjugated Dirac matter spinor $\Psi_E^C$. These are defined as

$$\Psi_E = \begin{bmatrix}\psi_+\\\bar{\psi}_-\end{bmatrix} \quad ; \quad \Psi_G = \begin{bmatrix}\sqrt{2}\tau\\\bar{\lambda}\end{bmatrix} \quad ; \quad \Psi_E^C = \begin{bmatrix}\psi_-\\\bar{\psi}_+\end{bmatrix} \ . \quad (19)$$

In terms of $\Psi_E$ and $\Psi_G$, the Lagrangian $\mathcal{L}$ will be

$$\mathcal{L} = \mathcal{L}_{KE} + \mathcal{L}_{gf} + \mathcal{L}_{m,M} + \mathcal{L}_{int} \ , \quad (20)$$



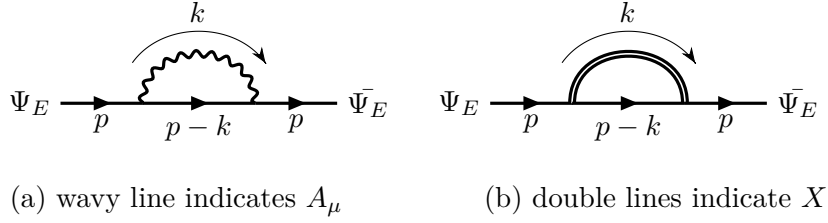

(a) wavy line indicates $A_\mu$     (b) double lines indicate $X$

FIG. 1. Electron self-energy diagrams

where

$$\mathcal{L}_{KE} = -\frac{1}{4}F^{\mu\nu}F_{\mu\nu} + \partial^\mu \phi_\pm \partial_\mu \phi_\pm^* + i\bar{\Psi}_E \gamma^\mu \partial_\mu \Psi_E + i\bar{\Psi}_G \gamma^\mu \partial_\mu \Psi_G$$
$$+\frac{1}{2}\partial^\mu X \partial_\mu X + \frac{1}{2}\partial^\mu Y \partial_\mu Y + i\lambda \sigma^\mu \partial_\mu \bar{\lambda} + 2i\tau \sigma^\mu \partial_\mu \bar{\tau} ,$$
$$\mathcal{L}_{gf} = -\frac{1}{2}(\partial_\mu \tilde{A}^\mu)^2 ,$$
$$\mathcal{L}_{m,M} = -m^2 \sum_{j=+,-} |\phi_j|^2 - m\bar{\Psi}_E \Psi_E + \frac{1}{2}M^2 \tilde{A}^\mu \tilde{A}_\mu - \frac{1}{2}M^2 X^2$$
$$-\frac{1}{2}M^2 Y^2 - M\bar{\Psi}_G \Psi_G ,$$
$$\mathcal{L}_{int} = iq(\tilde{A}_\mu + \frac{1}{M}\partial_\mu X) \sum_{j=+,-} j \left(\phi_j^* \partial_\mu \phi_j - \phi_j \partial_\mu \phi_j^*\right) - \frac{1}{2}q^2(|\phi_+|^2 - |\phi_-|^2)^2$$
$$+q(\tilde{A}_\mu + \frac{1}{M}\partial_\mu X) \left(q(\tilde{A}^\mu + \frac{1}{M}\partial^\mu X) \sum_{j=+,-} |\phi_j|^2 + \bar{\Psi}_E \gamma^\mu \Psi_E\right)$$
$$+\frac{iq}{\sqrt{2}} \left(\phi_+^\dagger \bar{\Psi}_G(1-\gamma_5)\Psi_E - \phi_+ \bar{\Psi}_E(1+\gamma_5)\Psi_G\right.$$
$$\left.-\phi_-^\dagger \bar{\Psi}_G(1-\gamma_5)\Psi_E^C + \phi_- \bar{\Psi}_E^C(1+\gamma_5)\Psi_G\right) + qMY(|\phi_+|^2 - |\phi_-|^2) . \qquad (21)$$

We can identify the three-point and four-point vertices in the interaction Lagrangian $\mathcal{L}_{int}$. In the next section, we will work out the leading order contribution to the self-energy of the electron $\Psi_E$.

### III. ELECTRON SELF ENERGY AT ONE LOOP

The relevant interaction from eq.(21) which contribute to electron self energy at one loop involving $\tilde{A}_\mu$ and Stuckelberg field $X$ are $q\bar{\Psi}_E \gamma^\mu \tilde{A}_\mu \Psi_E$ and $\frac{q}{M}\bar{\Psi}_E \gamma^\mu (\partial_\mu X)\Psi_E$. These terms are depicted in Fig.1(a),(b). There are also other terms, similar to Yukawa interactions, contributing to the self-energy of electron (see Fig.2).



We will now elaborate the computation of the self-energy correction from the Feynman diagrams drawn in Fig.1(a) and Fig.1(b) respectively. At $\mathcal{O}(q^2)$, the self energy contribution $T_1(p,p)$ for the electron with momentum $p$ will be sum of the contributions from both the Feynman diagrams:

$$T_1(p,p) = T_{1a}(p,p) + T_{1b}(p,p) \ . \tag{22}$$

Recall, $T_{1a}(p,p)$ is very similar to the standard three point QED vertex but with a massive photon $A_\mu$. The contribution $T_{1b}(p,p)$ for the Fig.1(b) involves the three point vertex with Stuckelberg component field $X$ in place of $A_\mu$. For the massive photon or Stuckelberg field, the self-energy corrections will be IR finite. However, we would like to work out the correction in the massless limit of the photon to be consistent with the experimental bound[25]. In the massless limit of the photon, we know that the longitudinal polarization of the photon disappears but the amplitude will become IR divergent. Hopefully the other diagrams involving Stuckelberg field and the superpartners should annul the IR divergence due to photon resulting in a IR finite contribution to the self energy of the electron.

Using the Feynman rules and the propagator for massive vector fields, the self energy for Fig.1(a) is

$$\begin{aligned} T_{1a}(p,p) &= \int \frac{d^4k}{(2\pi)^4} \frac{-id_{\mu\nu}(iq\gamma^\mu)i(\not{p}-\not{k}+m)(iq\gamma^\nu)}{(k^2-M^2+i\epsilon)([p-k]^2-m^2+i\epsilon)} \\ &= -q^2 \int \frac{d^4k}{(2\pi)^4} \frac{d_{\mu\nu}\gamma^\mu(\not{p}-\not{k}+m)\gamma^\nu}{(k^2-M^2+i\epsilon)([p-k]^2-m^2+i\epsilon)} \ . \end{aligned} \tag{23}$$

Recall the propagator for the virtual massive photon is

$$\frac{-i(d_{\mu\nu})}{k^2-M^2} \ ,$$

where $d_{\mu\nu}$ is

$$d_{\mu\nu} = \eta_{\mu\nu} - \frac{1}{M^2}k_\mu k_\nu = -\sum_{\lambda=1}^{4}\epsilon^*_\mu(k,\lambda)\epsilon_\nu(k,\lambda) \ , \tag{24}$$

where $\epsilon(k,\lambda)$ is the polarization vector for photon momentum $k$ and helicity $\lambda$. In the massless limit, the polarisation vector $\epsilon_\mu(k,\lambda)$ can be approximated as

$$\epsilon_\mu(k,\lambda) = \frac{k_\mu}{M} + \mathcal{O}\left(\frac{M}{k}\right) + .... \ , \tag{25}$$

which implies

$$\lim_{M \to 0} d_{\mu\nu} \approx -\frac{k_\mu k_\nu}{M^2} \ .$$



Substituting the above $g_{\mu\nu}$ in eqn.(23), we have the following divergent term $\propto \frac{1}{M^2}$ for $T_{1a}(p,p)$:

$$T_{1a}(p,p) = \frac{q^2}{M^2} \int \frac{d^4k}{(2\pi)^4} \frac{\slashed{k}(\slashed{p}-\slashed{k}+m)\slashed{k}}{(k^2 - M^2 + i\epsilon)([p-k]^2 - m^2 + i\epsilon)} \ . \tag{26}$$

The above expression can be interpreted as a IR divergent term when photon mass $M \to 0$.

Using the Feynman rules and propagator for the massive Stuckelberg scalar component $X$, the self energy $T_{2b}(p,p)$ for Fig.1(b) is

$$\begin{aligned}T_{2a}(p,p) &= \frac{q^2}{M^2} \int \frac{d^4k}{(2\pi)^4} \frac{i(-k_\mu)\gamma^\mu i(\slashed{p}-\slashed{k}+m)(-k_\nu)\gamma^\nu}{(k^2 - M^2 + i\epsilon)([p-k]^2 - m^2 + i\epsilon)} \\ &= -\frac{q^2}{M^2} \int \frac{d^4k}{(2\pi)^4} \frac{\slashed{k}(\slashed{p}-\slashed{k}+m)\slashed{k}}{(k^2 - M^2 + i\epsilon)([p-k]^2 - m^2 + i\epsilon)} \ . \end{aligned} \tag{27}$$

Note that the integrand in the above expression is equal and opposite to the integrand in the IR divergent eqn.(27). That is,

$$\lim_{M \to 0} T_1(p,p) = T_{1a}(p,p) + T_{1b}(p,p) = 0 \ . \tag{28}$$

Hence, the IR divergence in the massless limit ($M \to 0$) of the photon is exactly cancelled by the self energy correction due to the massless limit ($M \to 0$) of the Stuckelberg scalar component $X$ at one loop. That is, self energy correction from the diagrams in Fig.1 are IR finite in the $M \to 0$ limit.

There are also Yukawa interaction 3-point vertex diagrams contributing to self energy of electron at one loop as shown in Fig.2. The relevant Yukawa interaction terms involving Dirac spinor $\Psi_E$ (electron) and the Dirac spinor $\Psi_G$ (composed of gaugino spinor and Stuckelberg spinor) from eqn.(21) are

$$\begin{aligned}\mathcal{L}_{Yukawa} &= -\frac{iq}{\sqrt{2}} \left( \phi_+ \bar{\Psi}_E(1+\gamma_5)\Psi_G - \phi_+^\dagger \bar{\Psi}_G(1-\gamma_5)\Psi_E - \phi_- \bar{\Psi}_E^C(1+\gamma_5)\Psi_G \right. \\ &\qquad \left. + \phi_-^\dagger \bar{\Psi}_G(1-\gamma_5)\Psi_E^C \right) \\ &= -\frac{iq}{\sqrt{2}} \left( \phi_+ \bar{\Psi}_E(1+\gamma_5)\Psi_G - \phi_+^\dagger \bar{\Psi}_G(1-\gamma_5)\Psi_E - \phi_- \bar{\Psi}_G^C(1+\gamma_5)\Psi_E \right. \\ &\qquad \left. - \phi_-^\dagger \bar{\Psi}_E(1-\gamma_5)\Psi_G^C \right) \ . \end{aligned} \tag{29}$$

where $\Psi_E^C = \hat{C}\bar{\Psi}_E^T$ with $\hat{C} = -i\gamma^0\gamma^2$ denoting the charge conjugation operator. Using the definition of charge conjugation operator $\hat{C}$, we convert a charge conjugated Dirac spinor $\Psi_E^C$ for electron field into a charge conjugated Dirac spinor $\Psi_G^C$ for gaugino and Stuckelberg spinor field.



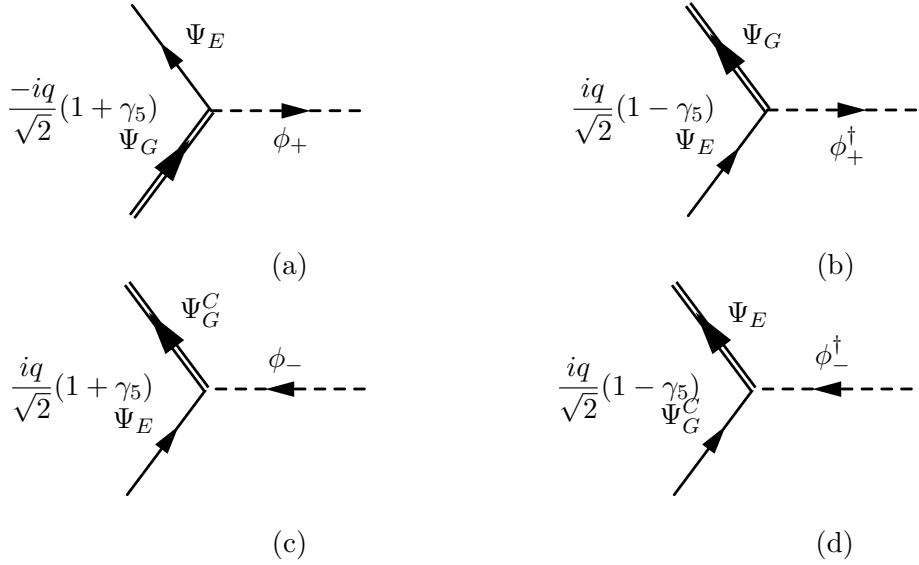

FIG. 2. Yukawa interaction vertex diagrams

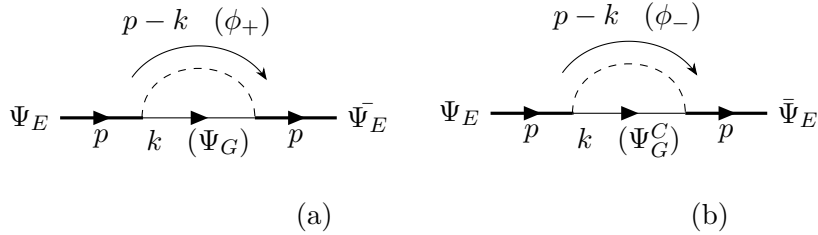

FIG. 3. Electron self-energy diagrams from Yukawa vertices

The interaction vertices for eqn.(29) are depicted in Fig.2 where solid lines represent electron Dirac spinor and solid double line represents the massive gaugino and Stuckelberg Dirac spinor. The dotted lines represent scalar fields $\phi_\pm$ which are superpartners of the electron.

From Fig.2, we choose the relevant Yukawa interaction diagrams with the electron field as the initial and final states to calculate self energy correction due to massive scalar field $\phi_\pm$.

We will now elaborate the computation of the self-energy correction $T_2(p,p)$ due to Yukawa interactions where $p$ is the momentum of the electron. At $\mathcal{O}(q^2)$, the electron self-energy correction $T_2(p,p)$ will be sum of the contributions from the Feynmann diagrams shown in Fig.3(a),(b):

$$T_2(p,p) = T_{2a}(p,p) + T_{2b}(p,p) \ . \tag{30}$$



Here, $T_{2a}(p,p)$ represents self energy correction due to massive scalar field $\phi_+$ and $T_{2b}(p,p)$ represents self energy correction due to massive scalar field $\phi_-$.

Using the Feynman rules and propagator for the massive scalar field, the self energy correction for Fig.3(a) is

$$T_{2a}(p,p) = \int \frac{d^4k}{(2\pi)^4} \frac{-iq(1+\gamma^5)(\slashed{k}+M)iq(1-\gamma^5)}{(k^2-M^2+i\epsilon)([p-k]^2-m^2+i\epsilon)}$$
$$= \frac{q^2}{2} \int \frac{d^4k}{(2\pi)^4} \frac{(1+\gamma^5)(\slashed{k}+M)(1-\gamma^5)}{(k^2-M^2+i\epsilon)([p-k]^2-m^2+i\epsilon)} \;, \qquad (31)$$

and for Fig.3(b) is

$$T_{2b}(p,p) = \int \frac{d^4k}{(2\pi)^4} \frac{iq(1+\gamma^5)(\slashed{k}+M)iq(1-\gamma^5)}{(k^2-M^2+i\epsilon)([p-k]^2-m^2+i\epsilon)}$$
$$= -\frac{q^2}{2} \int \frac{d^4k}{(2\pi)^4} \frac{(1+\gamma^5)(\slashed{k}+M)(1-\gamma^5)}{(k^2-M^2+i\epsilon)([p-k]^2-m^2+i\epsilon)} \;. \qquad (32)$$

The two eqns.(31) and (32) are exactly same differing in sign. Hence they cancel each other at $\mathcal{O}(q^2)$.

So far, we have elaborated on the relevant Feynman diagrams contributing to self-energy of the electron at the leading order $\mathcal{O}(q^2)$. We observe that the self energy of the electron in the Stuckelberg SUSY QED is free from infrared divergences in the massless limit $(M \to 0)$ of the photon and the Stuckelberg field. For completeness, we will focus on the electron-electron-photon vertex correction in the following section to investigate the IR divergences in the massless limit of the photon.

## IV. CANCELLATION OF IR DIVERGENCES IN VERTEX CORRECTION AT ONE LOOP LEVEL

The relevant 1PI Feynman diagrams at one loop which contribute to the electron-electron-photon vertex correction $\delta\Gamma^\mu(p',p)$ are shown in Fig.4(a) and Fig.4(b) where $p, p'$ are the momentum of the two external electrons. Hence vertex correction will be sum of the vertex corrections from the two diagrams:

$$\delta\Gamma^\mu(p',p) = \delta\Gamma_a^\mu(p',p) + \delta\Gamma_b^\mu(p',p) \;. \qquad (33)$$

Note that the vertex correction $\delta\Gamma_a^\mu(p',p)$ is the familiar QED vertex correction at one loop but with a massive photon $A_\mu$. The contribution $\delta\Gamma_b^\mu(p',p)$ for Fig.4(b) involves the three



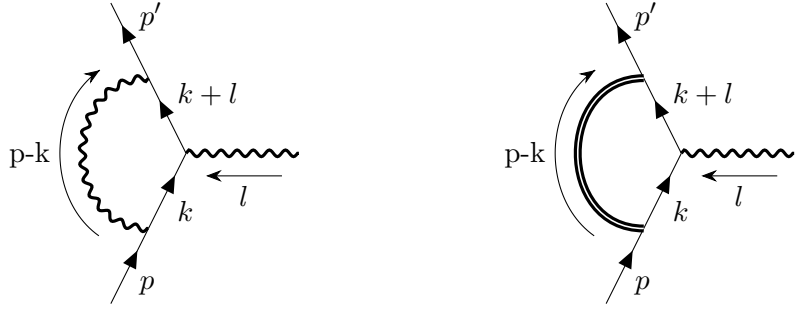

(a) wavy line indicates $A_\mu$    (b) double lines indicate $X$

FIG. 4. Electron vertex correction diagram

point vertex with Stuckelberg scalar component field $X$ replacing $A_\mu$ as an internal line in the loop. We will now explicitly compute the vertex corrections corresponding to the Feynman diagrams drawn in Fig.4 and check whether the total contribution continues to remain IR finite in the massless limit $(M \to 0)$ of the photon and Stuckelberg field. Using Feynman rules and propagators for massive vector fields, the vertex correction for Fig.4(a) is

$$\delta\Gamma_a^\mu(p',p) = \int \frac{d^4k}{(2\pi)^4} \frac{-id_{\nu\rho}(iq\gamma^\nu)i(\slashed{k}+\slashed{l}+m)iq\gamma^\mu i(\slashed{k}+m)(iq\gamma^\rho)}{([k+l]^2-m^2+i\epsilon)(k^2-m^2+i\epsilon)([p-k]^2-M^2+i\epsilon)} \ . \tag{34}$$

Our interest is to work out the contribution when the photon mass $M \to 0$ which will give IR divergence. As mentioned in the previous section on self energy correction, we will substitute $-d_{\nu\rho} = \frac{k_\nu k_\rho}{M^2}$ in the massless limit leading to

$$\delta\Gamma_a^\mu(p',p) = \frac{iq^3}{M^2} \int \frac{d^4k}{(2\pi)^4} \frac{\slashed{k}(\slashed{k}+\slashed{l}+m)(\slashed{k}+m)\slashed{k}}{([k+l]^2-m^2+i\epsilon)(k^2-m^2+i\epsilon)([p-k]^2-M^2+i\epsilon)} \ . \tag{35}$$

Now, using the Feynman rules and propagator for the massive Stuckelberg scalar component $X$, the vertex correction corresponding to the Fig.4(b) is

$$\delta\Gamma_b^\mu(p',p) = -\frac{iq^3}{M^2} \int \frac{d^4k}{(2\pi)^4} \frac{\slashed{k}(\slashed{k}+\slashed{l}+m)(\slashed{k}+m)\slashed{k}}{([k+l]^2-m^2+i\epsilon)(k^2-m^2+i\epsilon)([p-k]^2-M^2+i\epsilon)} \ . \tag{36}$$

Clearly, the integrals (35, 36) are equal and opposite cancelling in the massless limit of $A_\mu$ field. Thus we see that one loop vertex correction is also free of IR divergences.

## V. CONCLUSIONS

It is well known that infrared question in QED can be resolved using practical approach of measuring instrument's ability to distinguishing soft photons[1–3]. It was also resolved by Faddeev and Kulish by invoking a new Hilbert space of coherent states of electrons and photons. Recently this led to new understanding in terms asymptotic symmetries [35, 36] in theories with massless representations.

The above theories did not incorporate the practical question of measuring mass of photon raised by Schrodinger[24]. Currently, the experimental upper bound of the photon mass turns out to be $\leq 10^{-18} eV$[25]. Note that the introduction of mass to photon, however small it is, along with maintaining gauge invariance introduces new degrees of freedom which have gravitational interactions. This led to the interest in the study of Stuckelberg QED[12] to verify IR finite QED results in the massless limit of photons. The ultralight Stuckelberg mass appears to be a plausible candidate for dark matter[31, 32]. In order to ensure that the particle is stable, we will require additional symmetry protecting mass renormalisation.

Supersymmetry (SUSY) is one such symmetry which was essential to avoid tachyonic modes in string theory. Further, SUSY was introduced in phenomenology to avert quadratic divergences in the self energy corrections of scalars. Keeping in mind the objective of obtaining stable dark matter ultralight candidates, the extensions of Stuckelberg QED with an additional $\mathcal{N}=1$ supersymmetry (Stuckelbergy SQED) may be one of the gauge theories to pursue in future.

We know that the $\mathcal{N}=1$ supersymmetry QED (SQED) has IR divergences. Hence, in this paper we briefly reviewed the Lagrangian describing Stuckelberg SQED and focussed on self energy and vertex correction. Particularly, we investigated the leading order self energy correction of electron as well as the electron-electron-photon vertex correction in the massless limit of photon $M$ and Stuckelberg field. Interestingly, we explicitly showed that the IR divergences are cancelled by superpartner diagrams reinforcing that the theory is IR finite at leading order $\mathcal{O}(q^2)$ in the $M \to 0$ limit.

We could have added Fayet illioupulos term too, but that would not have changed the situation. Verifying the extension to all loops, though is definitely of academic interest and will be attempted in future. The theory with supersymmetric extension we studied provides more additional degrees of freedom $X, \tau$. Further, these fields do not interact with matter[28, 29] but will have only gravitational interactions. The cosmological consequences would be interestng which will also be taken up for further analysis.

**Acknowledgements:** TRG would like to thank IIT Bombay for visit when the details of work were carried out. TRG is grateful to Hermann Nicolai, AEI, MPG, Potsdam for support where this work got initiated. RV would like to thank DST-INSPIRE for their financial support. RV would like to acknowledge the fruitful discussions with Sreerup Raychaudhuri. The authors would like to acknowledge Jai More and Deepesh Bhamre for their contribution when the work was initiated.

---